\documentclass{ws-ijmpcs}
\usepackage{amsmath}
\usepackage{amsfonts}
\usepackage{cancel}
\usepackage{lmodern,dsfont}
\usepackage{amssymb}
\newcommand{\be}{\begin{equation}}
\newcommand{\ee}{\end{equation}}
\newcommand{\ba}{\begin{align}}
\newcommand{\ea}{\end{align}}

\begin{document}

%\markboth{A. Mart\'inez Torres, M. Bayar, D. Jido and E. Oset}
%{STUDY OF THE $\bar K N$ SYSTEM AND COUPLED CHANNELS IN A FINITE VOLUME}

\catchline{}{}{}{}{}
%
%%%%%%%%%%%%%%%%%%%%%%%%%%%%%%%%%%%%%%%%%%%%%%%%%%%%%%%%%%%%%%%%%%%%

\title{\bf{STUDY OF THE $\bar K N$ SYSTEM AND COUPLED CHANNELS IN A FINITE VOLUME}}

\author{A. MART\'INEZ TORRES}

\address{Instituto de F\'isica, Universidade de S\~ao Paulo, C.P. 66318, 05389-970 S\~ao Paulo, SP, Brazil.}

\author{M. BAYAR}

\address{Department of Physics, Kocaeli University, 41380, Izmit, Turkey}

\author{D. JIDO}

\address{Department of Physics, Tokyo Metropolitan University, Minami-Ohsawa 1-1, Hachioji-shi, 192-0397 Tokyo, Japan}

\author{E. OSET}

\address{Departamento de F\'isica Te\'orica and IFIC, Centro Mixto Universidad de Valencia-CSIC Institutos de Investigaci\'on de Paterna, Aptdo. 22085, 46071 Valencia, Spain}

\maketitle

\begin{history}
\received{Day Month Year}
\revised{Day Month Year}
\end{history}

\begin{abstract}
We investigate the $\bar KN$ and coupled channels system in a finite volume and study the properties of the $\Lambda(1405)$ resonance.  We calculate the energy levels in a finite volume and solve the inverse problem of determining the resonance position in the infinite volume. We devise the best strategy of analysis to obtain the two poles of the $\Lambda(1405)$ in the infinite volume case, with sufficient precision to distinguish them. 
\end{abstract}

\ccode{PACS numbers: 11.25.Hf, 123.1K}

\section{Introduction}	
One of the challenges of lattice QCD calculations is to determine the spectra of mesons and baryons. However, in the case of resonances the main difficulty is that they do not correspond to isolated energy levels in the spectrum of the QCD Hamiltonian on the lattice.  In the case of one or two channels, the problem is formally solved under the framework of  L\"uscher~\cite{lu1,lage}, which relates the measured discrete value of the energy in a finite volume to the scattering phase shift at the same energy, for the same system in the infinite volume.

In the last decade, two and three hadron systems have been studied extensively with unitary chiral models, $U\chi PT$,  and dynamical generation of many baryon and meson resonances has been found \cite{bernard,osra,oller,mko1,mko2,mko3}. Recently, the unitary chiral models have also been extended to investigate the interaction of two hadrons in a finite volume~\cite{mi1,koren}. Within this method it is possible to determine the energy levels of the system considered in a box and from there determine physical observables in infinite volume, providing in this way predictions and analyzing methods which can be very helpful for Lattice QCD studies.  

In the present case we study the $\bar K N$ system and coupled channels, in which the  $\Lambda(1405)$ state is formed~\cite{osra}, in a finite volume following the model developed in Ref.~\refcite{mtb}. 
\section{Formalism}
\subsection{Scattering matrix in infinite volume}
The scattering matrix in infinite volume in the $U\chi PT$ is obtained by solving the Bethe-Salpeter equation 
\begin{equation}
T(E)=[1-V(E)G(E)]^{-1}V(E).\label{bse}
\end{equation}
In Eq.~(\ref{bse}), $E$ is the energy of the system in the center of mass frame, $V$ represents the matrix for the transition potentials  between the channels and $G$  is the meson-baryon loop function, given by:

\begin{equation}
G=\int \frac{d^3 q}{(2\pi)^3}\frac{M}{E_1 E_2}\frac{E_1+E_2}{E^2-(E_1+E_2)^2+i\epsilon},\label{loop}
\end{equation}
with $M$ the baryon mass and $E_1$, $E_2$ the energies of the particles propagating. This loop function is divergent and needs to be regularized with a cut-off or using dimensional regularization~\cite{osra,oller}.  

\subsection{Energy levels in a finite volume}
To calculate the eigenenergies in a finite box of volume $V$, we just need to replace in Eq.~(\ref{loop}) 
\begin{align}
\int \frac{d^3 q}{(2\pi)^3}\to\frac{1}{V}\sum_{\vec{q}_L},\label{gbox}
\end{align}
which makes that $G\to \tilde{G}$, and look for the poles of the scattering matrix. That is, to solve the equation
\begin{equation}
\textrm{det}(1-V\tilde{G})=0\label{det}.
\end{equation}
If the box considered is symmetric, then $V=L^3$, with $L$ the side length, and imposing periodic boundary conditions we have that $\vec{q}_L=\frac{2\pi}{L}\vec{n}$, with $n\in \mathds{Z}^3$. In case of an asymmetric box, $V=L_xL_yL_z$ and $\vec{q}_L=2\pi\left(\frac{n_x}{L_x},\frac{n_y}{L_y},\frac{n_z}{L_z}\right)$. We will also consider the case where the meson-baryon system moves with a four momentum $P=(P^0, \vec P)$ in the box. In this case we follow the approach of Ref.~\refcite{mishamoving} and use the boost transformation from the moving frame to the center of mass frame.
\subsection{The inverse problem}
In our formalism, we can simulate ``synthetic" lattice data considering points related to the energy levels obtained in the finite volume and assigning to them a typical error of $\pm$ 10 MeV. To solve the inverse problem, i.e, getting the poles of the $\Lambda(1405)$ from the  ``synthetic" lattice data, we consider a potential with the same energy dependence than the chiral potential used to generate the energy levels in the finite volume, that is,
\begin{equation}
V_{ij}=a_{ij}+b_{ij}[E-(m_K+M_N)],\label{Vpara}
\end{equation}
but treat $a_{ij}$ and $b_{ij}$ as parameters which are determined by fitting the corresponding solutions for the energy levels to the ``synthetic" lattice data considered. 
\section{Results}
\subsection{Periodic boundary conditions in a symmetric  box}
\begin{figure}
\centering
\includegraphics[width=0.45\textwidth]{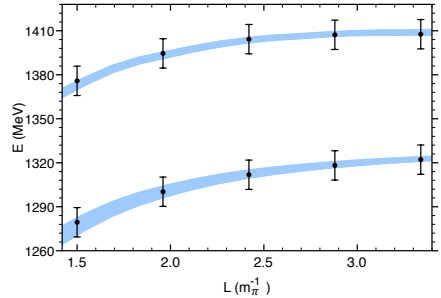}
\caption{(Color online) First two energy levels as function of the box side length $L$ for a symmetric box. The band corresponds to different choices of parameters within errors.}\label{ressym1}
\end{figure}

As an example, we show in Fig.~\ref{ressym1} the results of the energy levels reconstructed from the best fits to the ``synthetic" lattice data obtained in a symmetric box with periodic boundary conditions. The shadowed band in the figure corresponds to the random choices of parameters satisfying the condition $\chi^2\leqslant\chi^2_{min}+1$.
Using the potentials obtained from the fit and the loop function $G$ in infinite volume, we can solve Eq.~(\ref{bse}) and calculate the two-body $T$ matrix in the unphysical sheet, which allows us to determine the pole position of the $\Lambda(1405)$ associated to the band of solutions shown in Fig.~\ref{ressym1}. As a result we get a double pole structure for the $\Lambda(1405)$, with one pole in the region 1385-1433 MeV and half width between 93-137 MeV (which we call pole 1) and another one in the energy region 1416-1427 MeV and half width in the range 11-20 MeV (which we call pole 2). 
  
\subsection{Periodic boundary conditions in an asymmetric  box}
We consider now the case of an asymmetric box of side lengths $L_x=L_y=L$ and $L_z=zL$ to solve the inverse problem. In this case,  we use 5 points for level 0 calculated with $z=2.5$, 10 points for level 1 (5 for the case $z=0.5$ and 5 more for $z=2.0$) and 5 points for level 2 obtained with $z=2.0$.

Using the solutions obtained from the fits to the ``synthetic" lattice data, we can solve the Bethe-Salpeter equation in an infinite volume. In this case we get a double pole structure for the $\Lambda(1405)$ with pole positions at $(1383-1407) -i (57-69)$ MeV and $(1425-1434)-i (25-35)$ MeV. 

\subsection{Periodic boundary conditions in a moving frame}
In this case, we consider levels 0 and 1 determined for 5 different values of the center of mass momentum and two points in each of these curves. Using this ``synthetic" lattice date, the solution of Eq.~(\ref{bse}) in this case shows two poles for the $\Lambda(1405)$: one at $(1388-1418) -i(59-77)$ MeV and other at $(1412-1427)-i(16-34)$ MeV.
\section{Conclusions}
If we compare the results found for the poles of the $\Lambda(1405)$ using the different data sets considered with the ones of the chiral model~\cite{osra}, $1390-i66$ MeV and $1426-i16$ MeV, respectively, we find that the cases of an asymmetric box and of a symmetric box but in a moving frame seem to be more suited to get the two poles of the $\Lambda(1405)$ with more precision.

\section*{Acknowledgments}
A.~M.~T  thanks the Brazilian funding agency FAPESP for the financial support. This work is partly supported by the Spanish Ministerio de Economia y Competividad and European FEDER fund under the contract number FIS2011-28853-C02-01 and the Generalitat Valenciana in the program Prometeo, 2009/090 and by the Grant for Scientific Research (No.~24105706 and No.~22540275) from 
MEXT of Japan. M. Bayar acknowledges support through TUBITAK, BIDEP-2219 grant. 

%\begin{thebibliography}{000} %for 3 digits
%\begin{thebibliography}{00}  %for 2 digits

\end{document}